\documentclass[a4paper,12pt]{article}
\usepackage{amsmath,amssymb}

\begin{document}

\begin{center}
\section*{A short study of  a string on a plane: the  energy and the effective mass}

{S.V. Talalov}

\vspace{5 mm}

{\small Department of Applied Mathematics, State University of Tolyatti, \\ 14 Belorusskaya str.,
 Tolyatti, Samara region, 445020 Russia.\\
svt\_19@mail.ru}

\end{center}

\begin{abstract}
We investigate the new special class  of the finite string on a plane, after the reduction from the relativistic $4D$ case.
The suggested special form of the phase space allows to define the extended Galilei group as a group of the space - time symmetry for the considered system.
The definition of the energy for the studied  non-relativistic string through the Cazimir function of this  group is
suggested.  The concept of the effective mass for the investigated dynamical system is introduced.
The appearance of strong correlations between the degrees of freedom even on the classical level is discussed.
\end{abstract}

{\bf keywords:} {non-critical strings;   planar quaziparticles, effective mass}

{PACS Nos.: 11.25.Pm, 12.90.+b}

\vspace{5mm}

 {\bf I.} 
 The idea that the point-like singularities of  one - dimensional structures in  certain crystals can be interpretted as  some quasiparticles is very old
 (see\cite{Andr_76}, for example). On the other hand, the unconventional points of view on elementary particles continue to appear. 
 So, one of them\cite{LevWen} describes    ''particles''  as  defects of the string condensed matter
(''string-net condensation'').  
Speaking about  quasiparticles on the plane, we take into account the importance of the  ''anyon''  studies. 
They  are a starting point for the topological quantum calculations.
Thus,  studies of  particle-like excitations of  planar infinite - dimensional dynamical systems are of a certain interest.
It seems that the string\cite{Zwi} is the  simplest mathematical model of this type of system.
 In the general case, a planar string has  cuspidal points that can be interpretted as a certain particle - like structures.
   There are numerous works devoted to the appearance and the dynamics of these points\cite{BruEls,KliNik,And}.
  However, our approach to the string dynamics   is  completely original. 
  The main difference is the non-standard hamiltonization of the considered theory. 
As was noted by Dirac many years ago\cite{Dirac}, the various hamiltonian descriptions  of one and the same dynamical system lead, in general, to the different physical results.
This idea is still actual: the related articles appear  from time to time (see\cite{Magri,MSSV}, for example).

  Certain specific models in this aspect  have been realised in the earliest 
 author's works\cite{Tal_IJMPh,Tal_Nova},  where the theory  was investigated on
 both classical and quantum levels. 
  Among other results, it was found that the world - lines of the  cuspidal points  form the non-trivial braids in the space-time. 
 
 In this article we consider  the new  special class of  non-relativistic planar  finite strings and suggest a non - canonical approach to the description of the dynamics.
  As a result,  we  suggest:    firstly,   an approach to  energy definition  and, secondly,
 the  concept of the effective mass for the  considered non-relativistic dynamical system.    
 As regards the possible interpretation, we assume that suggested model could be useful for the analysis of some one-dimensional structures on a plane.  
 Examples here could be long polymer molecules, the boundaries between the two continuum media  and so on.
 A separate direction here is the application to development of the string - net hypothesis that was mentioned above (see, also\cite{FFNWW}). Indeed, the planar string is the 
 ''elementary atom''  for this structure.
 
 Let us recall the main points of our approach to the description of the planar string dynamics. The details and proofs can be found in the works cited above.
 Certain formulae will be modified in accordance with our purposes.
  \begin{enumerate}
  \item Believing in four-dimensionality of the world, 
     we start from the  Nambu-Goto string    in Minkowski space-time $E_{1,3}$.  This object is described by the radius-vector ${\bf X} = {\bf X}(\xi^0,\xi^1)$ where the 
    dimensionless parameters $\xi^0$ and   $\xi^1$ describe the    evolution and space parametrization of the  string. 
   We use the well-known orthonormal gauge so that the equations $\partial_{+}\partial_{-}{\bf X} =0$
   and the ''constraints'' $ (\partial_{\pm}{\bf X})^2 = 0$ hold. The symbols $\partial_\pm$ denote the derivatives $\partial/\partial \xi_\pm $ where the cone parameters $\xi_\pm =\xi^1\pm\xi^0$.
   Apart from the equalities, this point gives  our theory the dimensional constant\footnote{The proportionality coefficient  in the string action.} $\gamma$ and the
    standard formulae for the dynamical invariants (energy-momentum and angular momenta).
  \item  The standard  boundary conditions   for the case $\xi^1\in [0,L]$ take place: 
\begin{equation}
  \label{bc_x}
  \partial_{1}{\bf X}\Bigg\vert_{\,\xi^1=0}= \quad   \partial_{1}{\bf X}\Bigg\vert_{\,\xi^1 =L} \quad =0\,.
   \end{equation}
    The direct consequense of the condition at the point $\xi^1=0$ will be the equality
    $ \partial_{+}{\bf X}(\xi) = -  \partial_{-}{\bf X}(-\xi)\,.$
    Therefore, we can consider the single vector- function  on the interval $[-L,L]$  
    \begin{equation}
    \label{dX-single}
    \partial{\bf X}(\xi) =\varepsilon \partial_\varepsilon{\bf X}(\varepsilon\cdot\xi)\,,
    \end{equation}
    where $\varepsilon =\varepsilon(\xi) \equiv sign(\xi)$,  instead of two functions $\partial_{\pm}{\bf X}$ on the interval $[0,L]$.
        Further we investigate the special case  of the finite  string so that the parameter $\xi^1\in [0,\infty)$. 
     The limit  $L\to\infty$ conserves the conditions (\ref{bc_x}), but the value $\lim_{L\to \infty} \partial_{0}{\bf X}$ is undefined.
   To ensure the convergence of improper integrals in the subsequent studies, we
   limit ourselves to the consideration  of the  special class of  string configurations. 
   We    suppose that the vector-function $\partial{\bf X}(\xi)$ has the form
   \begin{equation}
 \label{e_pm}
 \partial{\bf X}(\xi)  =  {\varkappa}\,\exp\Bigl(-{p}\,(\xi + q)^2\Bigr) {\bf e}(\xi)\,,
  \end{equation}
  where $p >0$, $q$  and  ${\varkappa}$  are definite constants and  the vector function ${\bf e}(\xi)$  has  bounded components.
   The zero component of this vector - field  must               satisfy the condition   $\lim_{\xi\to -\infty} e_0(\xi) = 1$.
    For every value of the parameter $\xi$ the vector   ${\bf e}(\xi)$ is a dimensionless light - like projective (scale - invariant) vector. 
      Thus the representation (\ref{e_pm}) separates out the scale-transformed mode  $\varkappa$.
             Speaking about the separation of the scale-transformed mode, this step is  quite natural in our opinion          
         because  the  Nambu-Goto action  describes the scale-invariant theory.
                  As regards our subsequent purposes here, we intend  to separate the general set of  dynamical variables into two groups.
         The first group (internal variables)   will be  invariant in relation to both the scale  transformations and motions of the target space, the second group
          (external variables) will be transformed     in some way.               
            \item  
       Let us select an orthonormal basis
  ${\bf e}_{\nu}(\xi)$  so that  
 ${\bf  e}=\left({\bf e_{0}} - {\bf e_{3}} \right)/2$,  ${\bf e}_{\mu}{\bf e}_{\nu} = g_{\mu\nu}$. Obviously, this basis exists but is non - unique.
     The arbitrariness which arises here will be taken  into account later. 
     After that we introduce the vector - matrices ${\bf\widehat E}$  and ${\bf\widehat E}_0$ by means of the equalities
     ${\bf\widehat E} = {\bf e}_{\nu}(\xi){\bf M}^\nu$ and  ${\bf\widehat E}_0 = {\bf b}_{\nu}{\bf M}^\nu$  where the four vectors
      ${\bf b_\mu}$  form a certain stationary orthonormal basis and the four matrices
     ${\bf M}_\mu$   form  the  ''vector''  with  components $({\boldsymbol{1}}, {\boldsymbol{\sigma_i}})\,$.
     The matrices ${\bf\widehat E}$  and ${\bf\widehat E}_0$ are connected by the $SL(2,C)$ - valued field $T(\xi)$  by means of equality
$${\bf\hat E}(\xi) = T(\xi){\bf\hat E_0}T^{+}(\xi)\,.$$  
      We  make the reduction to $D = 1+2$ space-time  
through the requirement  $T \in SL(2,R)$. As a consequence,  
${\bf e}_{2}(\xi) \equiv  {\bf b}_2\,.$
Thus  the reduced  space-time is any space $E_{1,2} \perp {\bf b}_2$. 
    In the space $E_{1,2}$, we reconstruct the tangent vectors  $\partial_{\pm}{\bf X}$ with help of the formulae
    (\ref{dX-single})  and (\ref{e_pm}), where
                  \begin{equation}
  {\bf  e}(\xi)
 = \frac{1}{2} \Bigl[ \left( t_{11}^2 + t_{12}^2 \right)
  \,{\bf b}_0 -
    2\left( \,t_{11}{t}_{12}\right)\, {\bf b}_1
   - \left(t_{11}^2 - t_{12}^2\right)\,{\bf b}_3 \Bigr] \,.
\label{dX}
 \end{equation}
     The functions $t_{ij} = t_{ij}(\xi)$ are the  elements of the matrix $T(\xi)$.  Obviously, this matrix   solves the linear problem
     \begin{equation}
     T^{\,\prime}(\xi)+ Q(\xi)T(\xi) = 0 \,,
     \label{spect1}
     \end{equation}
where the coefficient matrices  $Q  = - T^{\,\prime}T^{-1}.$
   The arbitrariness that was mentioned above makes it possible
to reduce the matrix $Q$ to the form  
$$Q(\xi) = -{\varrho}\,(\xi)\boldsymbol{\sigma_+} + {\varrho}\,(\xi)\boldsymbol{\sigma_-}\,.$$ 
The function ${\varrho}\,(\xi)$ is  real - valued, therefore $T(\xi)\in SO(2)$  after this reduction.
The introduction of this function is justified from the geometrical viewpoint:
the coefficients of the second  fundamental form of the string world - sheet will be proportional to the functions ${\varrho}\,(\pm\xi_\pm)$.
In addition, it must be emphasised that  the executed  procedure $SL(2,R)\ni T \to T\in SO(2)$
does not commute with  the Lorentz boosts in the space - time $E_{1,2}$.  
Thus,  our further  theory is non-relativistic. 
These questions have  been investigated in the cited works in detail. 
\end{enumerate}

 {\bf II.}  Let us fix the type of the functions  $\varrho\,(\xi)$.
 To satisfy  our  requirement   $|{\bf  e}(\xi)| < const$,
we  make the following statement:
$$\varrho\,(\xi) = \rho(\xi) +\frac{1}{2}\,\omega\,, \qquad   \xi  \in(-\infty,\infty)   $$
where the real function  $\rho(\xi)\in  {\mathcal S}$ (Schwarz space) and the  real constant $\omega$ take  arbitrary values.
By virtue of these assumptions, the matrix  $T(\xi)$ that  solves the  system (\ref{spect1})
has the following form:
 \begin{equation}
\label{matr_T}
 {T}(\xi) =
 \left(\begin{matrix}
   \cos [\,I(\xi)+ \omega\xi/2+\beta\,],&  \sin [\,I(\xi)+ \omega\xi/2+\beta\,]\\
- \sin [\,I(\xi)+ \omega\xi/2+\beta\,],& \cos [\,I(\xi)+ \omega\xi/2+\beta\,]
\end{matrix}\right)\,,
  \end{equation}
where $I(\xi) = \int_{-\infty}^\xi\rho(\eta)d\eta\,$ and $\beta\in[0,2\pi)$. 
Thus,  we have a certain set of  independent dynamical variables that parametrize the string 
${\bf X}(\xi^0,\xi^1) = X_1{\bf b}_1 + X_3{\bf b}_3$ at this (intermediate)  stage of our investigations.
 This set consists of the scale - invariant quantities 
$\rho(\xi)$,  $\omega$,    $\beta$, $p$, $q$    and the scale - transformed quantities $\varkappa$ and  ${\bf Z}$.
The latter   are  the integration constants   required  for the reconstruction $\partial_{\pm}{\bf X} \to {\bf X}$. 
The group $E(2)$ transforms\footnote{The group of the motions of the plane $E_2$.} the variable $\beta$ and the constant vector ${\bf Z}$ only.
 We emphasize that   the identity $\varrho\,(\xi) \equiv 0$ should be fulfilled on any interval
 $[a,b]\subset (-\infty,\infty)$. This very fact makes it  impossible to trivialize the second fundamental form   by means of 
 certain conformal transformations
$\xi_\pm \to \widetilde\xi_\pm = A_\pm(\xi_\pm)\,$,  where $ A_\pm^{\prime} > 0$.

As a  result, we have the following expression for the complex - valued function ${\mathcal X} = X_3 +iX_1$:
\begin{eqnarray}
~&~&{\mathcal X}(\xi^0,\xi^1) = {\mathcal Z} - \frac{\varkappa}{2}\int\limits_{-\infty}^{\infty}d\eta
\Bigl[\varepsilon(\xi_+ -\eta) - \varepsilon(\xi_- +\eta)\Bigr]\times\nonumber\\
&\times& \exp\!\!\left[-{p}\,(\eta +q)^2 + i\Bigl(2I(\eta) +\omega\eta + 2\beta\Bigr)\right]\,, \quad \qquad
{\mathcal Z} = Z_3 + iZ_1\,.  \nonumber
\end{eqnarray}
Note  that the scale - invariant part of the function ${\mathcal X}(\xi^0,\xi^1)$   has the form of the Gabor transform for the function 
$\Bigl[\varepsilon(\xi_+ -\eta) - \varepsilon(\xi_- +\eta)\Bigr]\exp\Bigl(2iI(\xi)\Bigr)$.
  Let us write the   N\"oether invariants for  the considered string:
$$P_\mu = \gamma \int_{0}^{\infty} \partial_0 X_\mu\, d\xi^1\,,\qquad
M_{\mu\nu} = \gamma \int_{0}^{\infty} \left(\partial_0 X_\mu
X_\nu  -  \partial_0 X_\nu  X_\mu\right) d\xi^1\,. $$ 
Of course, we can use these formulae for the spatial indices only, because our reduction is non - relativistic.
The formulae (\ref{dX-single}),  (\ref{e_pm}),  (\ref{dX})  and (\ref{matr_T}) make it possible to calculate the components $P_\mu$ and   $M_{\mu\nu}$
 through the variables  introduced above; in accordance with our assumptions, all integrals converge.
The result  for the momentum  $P_3 + i P_1$, the    square of the momentum
${\bf P}^2 = P^2_1 + P^2_3$ and  the internal angular moment ${\sf S} = M_{13} - Z_1 P_3 - Z_3 P_1\,$ 
has the form:
\begin{equation}
\label{mass-spin}
P_3 + i P_1 = \gamma\varkappa{\mathcal I}_0(\omega)\,,\qquad
{\bf P}^2 ={\gamma^2\varkappa^2}{\mathfrak I}_P(\omega) \,,\qquad {\sf S} = 
({\gamma\varkappa^2}/2){\mathfrak I}_S(\omega)\,,
\end{equation} 
where 
\begin{eqnarray}
~&~&{\mathcal I}_0(\omega)      = \frac{1}{2} 
 \int\limits_{-\infty}^{\infty}d\xi \exp\!\!\left[-{p}\,(\xi +q)^2 + i\Bigl(2I(\xi) +\omega\xi + 2\beta\Bigr)\right]\,,\nonumber\\
\label{IS}
~&~&g(\omega)\equiv {\mathfrak I}_P(\omega)  -  i{\mathfrak I}_S(\omega)      =   \frac{1}{4}\!\iint\limits_{-\infty}^{\infty}\!\!
d\xi d{\zeta} \exp\!\!\left[{-{p}\,\Bigl((\xi+q)^2 + ({\zeta}+q)^2\Bigr)}\right]\times \nonumber\\ 
~~&\times& \left[\cos\Biggl(\!2\!\int_{\zeta}^\xi\!\rho(\eta)d\eta  +\omega(\xi - \zeta)\!\!\Biggr)
+  i\varepsilon(\xi - {\zeta})\sin\Biggl(\!2\!\int_{\zeta}^\xi \rho(\eta)d\eta +\omega(\xi - \zeta)\!\Biggr)\right].
\end{eqnarray}
To make the formulae shorter, we use the complex - valued functions here.
Note also  that ${\mathfrak I}_P(\omega) = |{\mathcal I}_0(\omega)|^2 >0$.
Let us exclude the  variable $\varkappa$ from the second and third equalities (\ref{mass-spin}). As result we have the equation that
connects the quantities ${\bf P}^2 = {\bf P}^2(\omega)$ and ${\sf S} = {\sf S}(\omega)$ for the same value of the variable $\omega$:
\begin{equation}
\label{P-S-lin}
 \Phi_1(\omega) \equiv  {\bf P}^2(\omega){{\mathfrak I}_S(\omega)} - 2\gamma  {\sf S}(\omega){{\mathfrak I}_P(\omega)}  = 0\,.
\end{equation}
The ''constraint'' $\Phi_1(\omega) = 0$ is not the only consequence of the equalities (\ref{mass-spin}).
Indeed, if we use the integral representation $\varepsilon(\eta) =(i/\pi)\int_{-\infty}^\infty(e^{-i\kappa\eta}/\kappa)\,d\kappa$,
we can write the  integral formula that connects   the quantities ${\bf P}^2(\omega)$ and ${\sf S}(\omega)$ for 
the different values of the variable $\omega$:
\begin{equation}
\label{P-S-int}
    \Phi_2(\omega) \equiv \,  {\sf S}(\omega) - \frac{1}{2\pi\gamma}\int\limits_{-\infty}^{\infty}\frac{{\bf P}^2(\kappa)}{\kappa - \omega}d\kappa   = 0\,.
\end{equation}
The  improper integral is understood here as  the Cauchy principal value. 
To deduce this equation we firstly   use
$$ \iint_{-\infty}^{\infty}\!d\xi d{\zeta}F(\xi,\zeta)\exp\left[{-{p}\,\Bigl((\xi+q)^2 + ({\zeta}+q)^2\Bigr)}\right]  =0 $$
for every antisymmetric function $F(\xi,\zeta)$ and, secondly, the identity $(d\varkappa/d\omega)\equiv 0$.
Combining equalities (\ref{P-S-lin}) and (\ref{P-S-int}) we can easily write the closed singular integral equations for both  the internal angular moment ${\sf S}$ and the  square of the momentum  ${\bf P}^2$. For example, the last value satisfies the equation
\begin{equation}
\label{P2-int}
{{\mathfrak I}_S(\omega)}{\bf P}^2(\omega) = 
\frac{{{\mathfrak I}_P(\omega)}}{\pi}\int\limits_{-\infty}^{\infty}\frac{{\bf P}^2(\kappa)}{\kappa - \omega }d\kappa\,.
\end{equation}
The methods for solving such equations are well-investigated \cite{Gakhov}.      In practice the expression
   ${\bf P}^2 ={\gamma^2\varkappa^2}{\mathfrak I}_P(\omega)$    gives the special solution of the equation (\ref{P2-int})   - such solution that depends on the single constant
   $\varkappa$. Can there be another solution of  the  equation (\ref{P2-int})?   As is known, the general solution of this equation contains the number of 
   constants $\varkappa_1,\dots,  \varkappa_n$, where the number $n = {\rm ind}{\sf G}$ is the index of the function   
      $ {\sf G}(\omega) =  -{g(\omega)}/{g^*(\omega)}\,$
      on a closed loop formed by the real $\omega$-axis and the point $\omega = \infty$.     Note that the function   $g(\omega)$        
      can be continued analytically  to the complex plane for $Im\,\omega >0$. Because  $Re\,g(\omega)>0$ for the real $\omega$  and  $g(\infty) = 0$, we have the equalities\cite{Gakhov}:
      $ {\rm ind}\,g  = - {\rm ind}\,g^* = 1/2$. Thus $n=1$ and the considered special solution of the  equation (\ref{P2-int}) is unique.
      For example, there is no solution of the equation (\ref{P2-int}) so that ${\bf P}^2(\omega) \equiv const$. This very fact is important to  us in our further investigations.

    {\bf III.} 	As noted above,   the  string energy  is not  defined for the non-relativistic reduction.        
In the works\cite{Tal_IJMPh,Tal_Nova} the  reduction of the space-time $E_{1,2}$ to the space and time $E_2\times {\sf R}$
was supplemented by the consideration of   the extended\,\footnote{We consider  
the  extension with  a single parameter $m_0$ only. The two - parameter extension of the Galilei group was applied for the anyon modelling in the work\cite{Jackiw}. }  Galilei group  instead of the $3D$ Poincare group.
After that  the Cazimir function ${\widehat C}_3 = \widehat H -  ({1}/{2m_0}){\widehat {\bf P}}^2$ 
for the extended Galilei algebra  was used to define the energy (the quantities $\widehat H$ and  ${\widehat {\bf P}}$   are the generators of the time and space translations
correspondingly).
Without going into all  details here, we note that the correct definition of the extended Galilei group as a group of the global symmetry was possible
because of a special selection of  independent dynamical variables. So, the variables  $P_1$, $P_3$ and ${\sf S}$ were declared as independent (global) variables instead of the variables
$\varkappa$ and $\beta$. The equality (\ref{P-S-lin}) was considered as a constraint.  Recall that  the results of the works\cite{Tal_IJMPh,Tal_Nova}
correspond to the case $\omega=0$   only.
Above we have shown that the dependence of 
 the  variables  $P_1$, $P_3$ and ${\sf S}$  on the  variable $\omega$ is not  trivial.
Therefore, the replacement $(\varkappa, \beta)\to (P_1,P_3;{\sf S})$ is possible locally only.
To perform  this procedure, we consider  the phase space   ${\mathcal H}$   as the product  bundle  ${\mathcal H} =    {\mathcal H}_2 \times {\mathcal B}$, where
  the fiber  ${\mathcal H}_{2}$  is the finite - dimensional  phase space  which corresponds to  some structureless free particle on a plane; the 
fundamental coordinates  here are the variables ${\sf S}$,  ${\bf P}$ and ${\bf B} =m_0[{\bf Z} - (\xi^0/\gamma){\bf P}]$.  
  The base  ${\mathcal B} = {\sf R}_\omega \times {\mathcal H}_{\rho}\times {\mathcal H}_{p,q}  $ is parametrized by the variables $\omega\in{\sf R}_\omega$,  $\rho(\xi) \in  {\mathcal H}_{\rho} $
  and  $(p,q)\in {\mathcal H}_{p,q}  $.    The extended Galilei group acts independently  on  each fiber ${\mathcal H}_{2}$ as a structural bundle group. 
  The equalities  (\ref{P-S-lin})  and (\ref{P-S-int})   define the indepenent constraints in the global phase space ${\mathcal H}$.
 From the geometrical viewpoint, the surface ${W}_\Phi$ of these constraints is 
  a  non - trivial  cross-section   of the bundle ${\mathcal H}$.  
 The  Poisson brackets are postulated in the form:
  \begin{eqnarray}
\{\rho(\xi),  \rho(\eta)\}  =   -\frac{\gamma}{4m_0{\sf E}_0}\, \delta^\prime(\xi - \eta)\,, \qquad
\{P_i,B_j\} = m_0\delta_{ij}\,,   \qquad  \{p,q\} = \frac{\gamma}{m_0{\sf E}_0}      \, \nonumber
\end{eqnarray}
(other possible brackets  equal  zero).   Note that  $\{ \Phi_1(\omega),\Phi_2(\overline\omega)\}\equiv 0$.
 The  value ${\sf E}_0$ defines the energy scale; together with constants $\gamma$ and $m_0$ it is the {\it in-put} dimensional constant
    in our theory\footnote{Because we reduced our theory from the relativistic case, we can avoid the introduction of the additional constant ${\sf E}_0$ assuming that
     ${\sf E}_0 = m_0c^2$.  Nevertheless, we conserve the constant ${\sf E}_0$ as the independent dimensional constant.}. 

    Because the Cazimir function    $C_3 $  is interpreted as the ''internal energy'' of a   dynamical system,        we can write the hamiltonian $H =H(\omega)$ in the  form:  
\begin{equation}
  H(\omega) = \frac{{\bf P}^2(\omega)}{2m_0} + h_{int}(\omega)  + \sum_i\ell_i\Phi_i(\omega) \nonumber   \,,
  \end{equation}
  where the functions $\ell_i$, $i=1,2$  are  Lagrange multipliers.
     The value  $h_{int}(\omega) $ is the energy that corresponds to the ''internal'' degrees of  freedom $(p,q)$,   $\omega$ and $\rho(\xi)$. 
  It would be natural if we define this quantity as the hamiltonian that generates the dynamics of the corresponding variables.
   Let the function   $f(\omega, {\sf S})$ be an arbitrary function of the annulators  $\omega$ and ${\sf S}$.  Taking into account   the definition of the Poisson brackets,   
    the  function
$$h_{int}(\omega)\equiv h_{int}[S,\omega;\,p,\,\rho(\xi)] =  {\sf E}_0\left( 2\int\rho^2(\xi)d\xi + p\right) + f(\omega, {\sf S})  $$
 provides the dynamics $ \rho(\xi^1)\to \rho(\xi^1+\xi^0) $  and $q \to q+ \xi^0$. Indeed,
  if we define the physical (i.e. dimensional) time $t$ as $t = (m_0/\gamma)\xi^0$, the canonical equations
 $${\partial\rho }/{\partial t} = \{h_{int},\rho  \}\,,\qquad    {\partial q }/{\partial t} =  \{h_{int},q \}\, $$
  take place (for $\ell_1 =0$). As regards the ''observables'' -- the functions of canonical variables -- the dynamics will be correct too.
  For example,  we have the following  representation  for the angle $\beta$:
  $$ \beta(p,q, \omega, \rho;  {\bf P}) = \frac{1}{2}\Bigl(\arg(P_3 + i P_1) - \arg{{\mathfrak I}_0[p,q,\omega,\beta;\rho]}\Big\vert_{\beta=0}\Bigr)\,,$$
  so that   $\beta \to \beta + \omega\xi^0/2$   and $ {\partial\beta }/{\partial t} =  \{h_{int},\beta \}$.
         What is the explicit form for the function $f(\omega, {\sf S})$?   Let us consider a special configuration  when $\rho(\xi) \equiv 0$ and $p\to 0$.   This case corresponds to the 
   straight rotating  (with the frequency  $\Omega =(\gamma/m_0)\omega$)   string with  the pulsating length, but with the constant  moment ${\sf S}$. 
   As it seems, the function  $f(\omega, {\sf S}) = \gamma\omega{\sf S}/2m_0$ 
   is adequate for the description of the energy here.  
   Therefore, the final expression for the hamiltonian  will be:
  \begin{equation}
  \label{energy_1}
   H(\omega) = \frac{{\bf P}^2(\omega)}{2m_0} + 
   {\sf E}_0\Biggl(2\!\int\limits_{-\infty}^\infty\!\rho^{\,2}(\xi)d\xi + p\Biggr)   + \frac{\gamma\omega{\sf S}(\omega)}{2m_0} + \sum_i\ell_i\Phi_i(\omega)   \,.
      \end{equation}
    As usual, we can substitute the  first type constraints after the calculations of  all Poisson brackets. Thus, taking into account 
    the formulae  (\ref{P-S-lin})  and   (\ref{P-S-int}), we write the expression for the energy ${\sf E}(\omega)$:
    \begin{equation}
  \label{energy_2}
   {\sf E}(\omega) = H(\omega)\Bigg\vert_{\,W_\Phi}   = \frac{{\bf P}^2(\omega)}{2m_{\sf eff}(\omega)} + 
   {\sf E}_0\Biggl( 2\!\int\limits_{-\infty}^\infty\!\rho^{\,2}(\xi)d\xi + p\Biggr)  \,,
      \end{equation}
    where the effective mass $m_{\sf eff}$ is defined as follows:
    $$\frac{1}{m_{\sf eff}(\omega)} \equiv \frac{\!\partial^2{\sf E}}{\,\,\partial |{\bf P}|^2}  =
    \frac{1}{m_0}\left[ 1 +\frac{\omega}{2}\frac{{\mathfrak I}_S(\omega)}{{\mathfrak I}_P(\omega)}\right]\,.$$
    Because the inequality ${\mathfrak I}_P(\omega) > 0$ is fulfilled, the value $m_{\sf eff}(\omega) \not = 0$, $\forall\,\omega\in(-\infty,\infty)$.
      On the contrary, the case when $m_{\sf eff}(\omega) \to \infty$ is possible because the 
      function ${\mathfrak F}(\omega) =  {\mathfrak I}_P(\omega) + (\omega/2) {\mathfrak I}_S(\omega) $ can have certain zeros $\varpi_m$.
    This fact means that  all modes of our dynamical system are strongly correlated when $\omega\to\varpi_m$.        
   The certain properties of the model will be specific near the points $\varpi_m$. For example, we have the linear dependence 
   ${\sf S} = \alpha {\sf E} + C$ when $\omega\not =\varpi_m$.   But the values ${\sf S}$  and    ${\sf E}$ become   independent when $\omega = \varpi_m$.
   All     these effects  require   studies on a quantum level, of course.
   Our experience in such studies\cite{Tal_IJMPh,Tal_Nova}  gives hope for some interesting results here.

  \end{document}